\documentclass[aps,prl,reprint,superscriptaddress,amsmath,amssymb]{revtex4-2}
\usepackage{graphicx}
\usepackage{bm}
\usepackage{todonotes} 
\usepackage{physics}

\usepackage{siunitx}
\usepackage[colorlinks=true, linkcolor=blue, citecolor=blue, urlcolor=blue]{hyperref}
\usepackage{siunitx}

\usepackage{braket}

\usepackage{color}
\usepackage{colordvi}  
\definecolor{Red}{rgb}{1,0,0}

\makeatletter
\def\authornote{\xdef\@thefnmark{$\dagger$}\@footnotetext}
\makeatother

\begin{document}

\title{Stimulated generation of indistinguishable single photons from a quantum ladder system}

\author{Friedrich Sbresny$^\dagger$}
 \affiliation{Walter Schottky Institut, Department of Electrical and Computer Engineering and MCQST, Technische Universit\"at M\"unchen, 85748 Garching, Germany}%
\author{Lukas Hanschke$^\dagger$}
 \email{lukas.hanschke@wsi.tum.de}
 \affiliation{Walter Schottky Institut, Department of Electrical and Computer Engineering and MCQST, Technische Universit\"at M\"unchen, 85748 Garching, Germany}%
 \authornote{These authors contributed equally.}
 \author{Eva Sch\"oll}
 \affiliation{Institute for Photonic Quantum Systems (PhoQS), Center for Optoelectronics and Photonics Paderborn (CeOPP) and Department of Physics, Paderborn University, 33098 Paderborn, Germany}%
\author{William Rauhaus}
 \affiliation{Walter Schottky Institut, Department of Electrical and Computer Engineering and MCQST, Technische Universit\"at M\"unchen, 85748 Garching, Germany}%
\author{Bianca Scaparra}
 \affiliation{Walter Schottky Institut, Department of Electrical and Computer Engineering and MCQST, Technische Universit\"at M\"unchen, 85748 Garching, Germany}%
 \author{Katarina Boos}
 \affiliation{Walter Schottky Institut, Department of Electrical and Computer Engineering and MCQST, Technische Universit\"at M\"unchen, 85748 Garching, Germany}%
\author{Hubert Riedl}
 \affiliation{Walter Schottky Institut, Department of Physics and MCQST, Technische Universit\"at M\"unchen, 85748 Garching, Germany}%
\author{Jonathan J. Finley}
 \affiliation{Walter Schottky Institut, Department of Physics and MCQST, Technische Universit\"at M\"unchen, 85748 Garching, Germany}%
\author{Klaus J\"ons}
 \affiliation{Institute for Photonic Quantum Systems (PhoQS), Center for Optoelectronics and Photonics Paderborn (CeOPP) and Department of Physics, Paderborn University, 33098 Paderborn, Germany}%
\author{Kai M\"uller}
 \affiliation{Walter Schottky Institut, Department of Electrical and Computer Engineering and MCQST, Technische Universit\"at M\"unchen, 85748 Garching, Germany}%

\date{\today}

\begin{abstract}
We propose a scheme for the generation of highly indistinguishable single photons using semiconductor quantum dots and demonstrate its performance and potential. The scheme is based on the resonant two-photon excitation of the biexciton followed by stimulation of the biexciton to selectively prepare an exciton. Quantum-optical simulations and experiments are in good agreement and show that the scheme provides significant advantages over previously demonstrated excitation methods. The two-photon excitation of the biexciton suppresses re-excitation and enables ultra-low multi-photon errors, while the precisely timed stimulation pulse results in very low timing jitter of the photons, and consequently, high indistinguishability. Since both control laser fields are detuned from the emission energy, the scheme does not require polarization filtering, facilitating high brightness approaching unity. Moreover, the polarization of the emitted single photons is controlled by the stimulation laser field, such that the polarization of the quantum light is deterministically programmable. 
\end{abstract}

\maketitle

Single photons are a key resource for applications in photonic quantum technologies, such as quantum key distribution~\cite{Bennett1984}, linear optical quantum computing~\cite{Knill2001, Kok2007} and Boson sampling~\cite{Aaronson2011, Aaronson2011-2}.
Over the past two decades, semiconductor quantum dots have played a pivotal role for deterministic single-photon sources.~\cite{Gazzano2013, Somaschi2016, Ding2016, Snijders2018, Wang2019, Uppu2020, Tomm2021}
This stems from their excellent optical properties, such as almost exclusive emission into the zero-phonon line, near-unity quantum efficiency, near transform-limited linewidth, high emission rates and ease of integration into photonic nanoresonators and waveguides to further enhance the emission rates and efficiency~\cite{Senellart2017, Trivedi2020}.
A wealth of different excitation schemes have been developed, such as non-resonant excitation, resonant excitation, two-photon excitation of the biexciton, phonon-assisted excitation, all of which have specific advantages and disadvantages~\cite{Trivedi2020}.
Resonant excitation techniques avoid the creation of free charge carriers that introduce noise in the electronic environment, enabling near transform-limited linewidth \cite{Prechtel2013, Kuhlmann2015}.
Furthermore, resonant excitation has been shown to produce excellent indistinguishability~\cite{He2013}, but the single-photon purity is limited to $g^{(2)}(0) \approx 10^{-2}$ due to re-excitation and subsequent emission of a second photon~\cite{Fischer2018a}.
This re-excitation can be strongly suppressed for two-photon excitation of the biexciton ~\cite{Hanschke2018,Schweickert2018}, however, the indistinguishability of the two emitted photons is then limited due to the cascaded emission~\cite{Scholl2020}.
Recently, acoustic-phonon mediated excitation has been proposed and confirmed in experiments as an attractive method to produce a high single-photon purity and indistinguishability comparable to resonant excitation without the need for polarization filtering~\cite{Reindl2019,Cosacchi2019, Gustin2020,Thomas2020}.
So far, an excitation scheme that combines both ultra-low multi-photon error rate, high indistinguishability and inherent polarization control has remained elusive.

In this letter we theoretically model and experimentally demonstrate a single-photon generation scheme that combines all the advantages of previously established excitation methods. The scheme is based on the resonant two-photon excitation of a biexciton~\cite{Brunner1994,Stufler2006} followed by timed stimulation to prepare the exciton. The two-photon excitation of the biexciton suppresses re-excitation and enables ultra-low multi-photon errors. The precisely timed stimulation pulse prepares the system in the exciton state at a predetermined time. This greatly reduces timing jitter in the preparation of the exciton state caused by the biexciton population lifetime, and consequently, reestablishes the high indistinguishability for photons emitted from the exciton to ground state decay. Since the emission energy is detuned from both driving laser fields no polarization filtering is required, which enables higher brightness than cross-polarized schemes. Moreover, the scheme allows to deterministically program the polarization of the emitted photon (H/V) via the polarization of the stimulation pulse~\cite{Santori2002a}. Experiments and simulations exploring the system dynamics are in very good agreement and confirm the validity of the theoretical model.

\begin{figure}
    \centering
    \includegraphics[width=\linewidth]{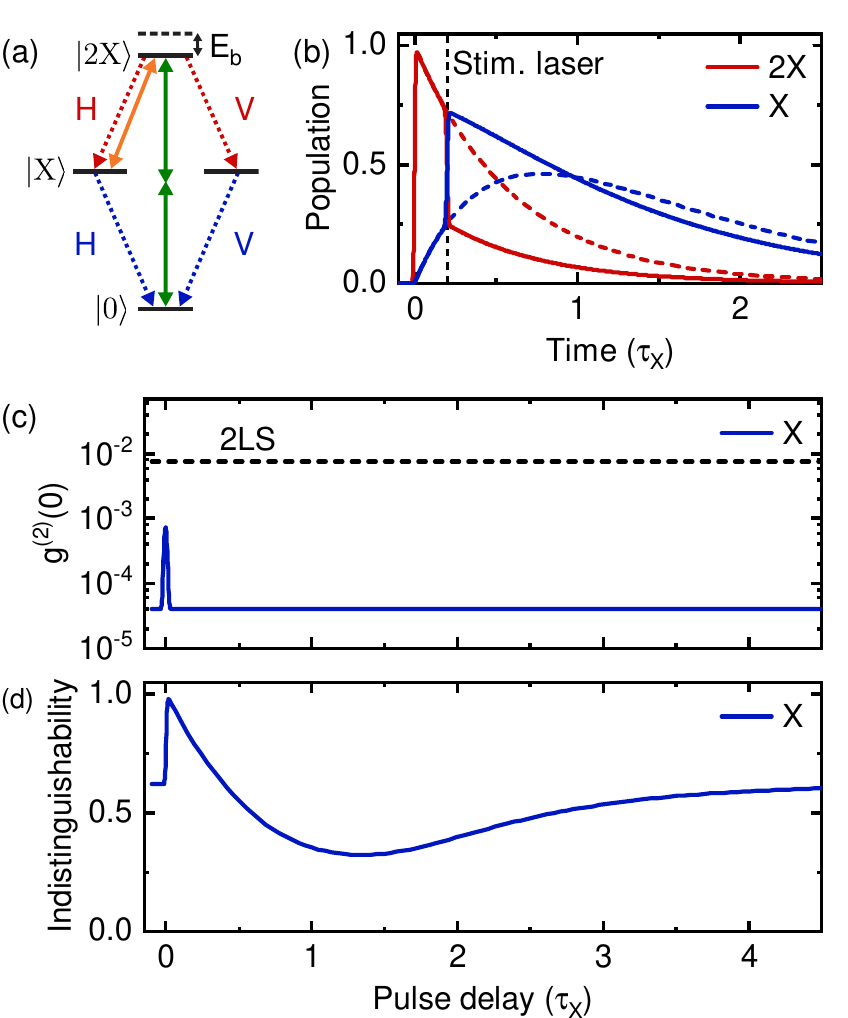}
    \caption{(a) Level scheme of the biexciton-exciton cascade: Solid arrows denote laser fields, dotted arrows denote emission through radiative decay. (b) Population evolution of the excited and intermediate state of a quantum ladder system excited via the two-photon resonance. The solid lines represent the state population with a stimulation laser pulse applied at $t=0.2\tau_{\mathrm{X}}$, the dashed lines represent the case  without stimulation laser. (c) Second-order coherence of the $\ket{\mathrm{X}} \rightarrow \ket{0}$ transition in dependence of the time delay between the excitation and stimulation pulse exhibits up to two orders of magnitude better $g^{(2)}(0)$ than for a resonantly driven two-level system (2LS) excited by a pulse of the same length (dashed line). (d) Numerically simulated indistinguishability in dependence of the pulse delay shows that for short pulse delays near-unity indistinguishability can be achieved for the emitted photons of the exciton decay.}
    \label{fig:figure1}
\end{figure}

The quantum level scheme of the system under study in this work is depicted in Fig.~\ref{fig:figure1}(a). It consists of the ground state $\ket{0}$, an excited state $\ket{\mathrm{2X}}$ and two intermediate states $\ket{\mathrm{X}}$ which are energetically split by a fine structure splitting~\cite{Bayer2002}.
The ground state $\ket{0}$ and the excited state $\ket{\mathrm{2X}}$ are resonantly coupled via a two-photon absorption process by a laser pulse (solid green arrows) resonant on a virtual level.
Note that due to the biexciton binding energy $E_b$, the $\ket{\mathrm{2X}} \leftrightarrow \ket{\mathrm{X}}$ transition energy is red-detuned by $E_b$ from the $\ket{\mathrm{X}} \leftrightarrow \ket{0}$ transition energy~\cite{Bayer1998}. 
The dotted arrows represent radiative decays. 
An additional linearly polarized stimulation laser pulse (solid orange arrow) resonant on the $\ket{\mathrm{2X}} \leftrightarrow \ket{\mathrm{X}}$ transition selectively couples the biexciton to one of the two exciton branches and thus the polarization of the pulse (H/V) determines the branch. 

To study the emission properties such as purity and indistinguishability of this system, we focus our attention on one of the two branches resulting in a three-level quantum ladder system.
The Hamiltonian of this ladder system in a reference frame oscillating at the excitation laser frequency then reads
\begin{equation}
\begin{split}
    H =& \frac{\left(\mu_{e} \cdot E_{e}(t) \right)^2 }{2E_b} \left(\ket{0} \bra{\mathrm{2X}}+ \ket{\mathrm{2X}}\bra{0} \right)\\ 
    &+ \frac{\mu_{s} \cdot E_{s}(t)}{2}\left( \ket{\mathrm{X}}\bra{\mathrm{2X}}+ \ket{\mathrm{2X}}\bra{\mathrm{X}}\right)
\end{split}
\end{equation}
where the first part is the excitation laser coupling between the ground and biexciton state after adiabatic elimination of the intermediate state.
The second term represents the coupling of the biexciton and exciton state through the stimulation laser. Here, the respective electric dipole moments of both transitions are denoted $\mu_{e,s}$ and the respective time-dependent electric fields of the applied laser fields are denoted $E_{e,s}(t)$.
Under pulsed excitation and stimulation, the system dynamics are dominated by coherent oscillations of the coupled states determined by the pulse area.
For the excitation pulse, the pulse area scales linearly with power
\begin{equation}
    A_{e}(t)= \int^{t}_{0} \dd{t'} \frac{\left(\mu_{e} \cdot E_{e}(t')  \right)^2}{\hbar E_b}
\end{equation}
while for the stimulation pulse the pulse area scales linearly with electric field 
\begin{equation}
    A_{s}(t)= \int^{t}_{0} \dd{t'} \frac{\left(\mu_{s} \cdot E_{s}(t')  \right)}{\hbar}
\end{equation}
We numerically model the system described above and its emission properties using the Quantum Toolbox in Python (QuTiP) and a quantum-optical master equation approach ~\cite{qutip1,qutip2,Fischer2016}.
The time evolution for the population of both excited states after excitation and subsequent stimulation is displayed in Fig.~\ref{fig:figure1}(b) in units of the exciton lifetime $\tau_{\mathrm{X}}$. 
At time $t=0$ the system is excited by a Gaussian $\pi$-pulse ($T_{\mathrm{pulse}} \ll \tau_{\mathrm{X}}$) resonantly at the ground state to biexciton transition. This leads to near-unity population of the $\ket{\mathrm{2X}}$ state (solid red line). 
In a free evolution period, the $\ket{\mathrm{2X}}$ state population decays leading to a build up of the $\ket{\mathrm{X}}$ state population (solid blue line). 
After an exemplary time delay $t=0.2 \tau_{\mathrm{X}}$ (dashed vertical line), the Gaussian stimulation pulse with an area of $\pi$ is applied which inverts the population of the biexciton and exciton state. 
The population evolution of the states without the stimulation pulse applied is depicted by the red and blue dashed lines, respectively. 
The time delay between the excitation and the stimulation laser pulse can be precisely varied, thus enabling direct manipulation of the effective emission rate $\gamma_{2X}$ only via optical means.
We set the parameters of the simulation, namely the emission rates of the two transitions and the pulse lengths of both pulses, in accordance with the values determined  for the InGaAs quantum dot experimentally investigated below.
By calculating the first- and second-order coherences $g^{(1)}(0)$ and $g^{(2)}(0)$ we determine the purity and indistinguishability of the single photons emitted from the $\ket{\mathrm{X}}\rightarrow \ket{0} $ transition~\cite{Fischer2016} as a function of the time delay between the two pulses.
The simulated degree of second-order coherence $g^{(2)}(0)$ is presented in Fig.~\ref{fig:figure1}(c) as a function of the time delay between the two laser pulses. 
Independent of the time delay, $g^{(2)}(0)$ is up to more than two orders of magnitude lower than for a resonantly driven two-level system characterized by the same emission rate and driven by a pulse with the same temporal shape (dashed line).
This results from the two-photon excitation which strongly suppresses re-excitation during the pulse duration \cite{Hanschke2018}.
At small time delays where both laser fields temporally overlap, $g^{(2)}(0)$ exhibits a sharp peak. When the pulses overlap, the decrease of the biexciton lifetime $\tau_{\mathrm{2X}}$ induced by the stimulation pulse increases the probability of re-excitation.
For larger time delays, the excitation protocol inherits the excellent single-photon purity from the two-photon excitation scheme as the stimulation occurs when the excitation drive is not present anymore. 

The major advantage of our proposed scheme is that it overcomes the existing inherent limits on the indistinguishability of photons emitted by a quantum ladder system.
It was found that for this type of system the indistinguishability of emitted photons depends on the lifetime ratio of both transitions, and for a single-photon source with negligible multi-photon emission is given by the relation~\cite{Simon2005,Scholl2020} 
\begin{equation}
    \mathbb{P}= \frac{\gamma_{2X}}{\gamma_{2X}+\gamma_{X}}
    \label{equ:equation3}
\end{equation}
which tends to unity for large $\gamma_{2X}$.
In Fig.~\ref{fig:figure1}(d) we present the numerically simulated indistinguishability as a function of the pulse delay. 
When the stimulation pulse arrives before the excitation pulse, the indistinguishability of the emitted photons is limited by Eq. (\ref{equ:equation3}) as it corresponds to the case for two-photon excitation only, which goes to $62.0\%$ for the chosen emission rates.
For short time delays between both pulses, the indistinguishability reaches a maximum of $97.9\%$ due to the increase of the biexciton emission rate arising from stimulation.
The maximum indistinguishability occurs where both pulses still have a very small overlap, causing $g^{(2)}(0)$ to be twice as large as for regular two-photon excitation. 
This could be further optimized by choosing shorter pulse lengths or different shapes, which would simultaneously suppress re-excitation and enable faster stimulated emission for the biexciton-exciton transition.
As the time delay between both laser pulses is increased, the indistinguishability decreases to a local minimum at ${\sim} 1.4\tau_{\mathrm{X}}$.
For this delay, the stimulation pulse effectively increases the excitation timing jitter, since population that had already decayed to $\ket{\mathrm{X}}$ is re-excited to $\ket{\mathrm{2X}}$.
For very long time delays ($t > 4 \tau_{\mathrm{X}} $), the indistinguishability approaches the accessible limit of indistinguishability for the two-photon excitation, since most of the population of the excited states has already decayed to the ground state before the stimulation pulse arrives.
Thus, by choosing a short time delay in between excitation and stimulation laser pulse our proposed scheme allows for the generation of highly indistinguishable single photons.

\begin{figure}
    \centering
    \includegraphics[width=\linewidth]{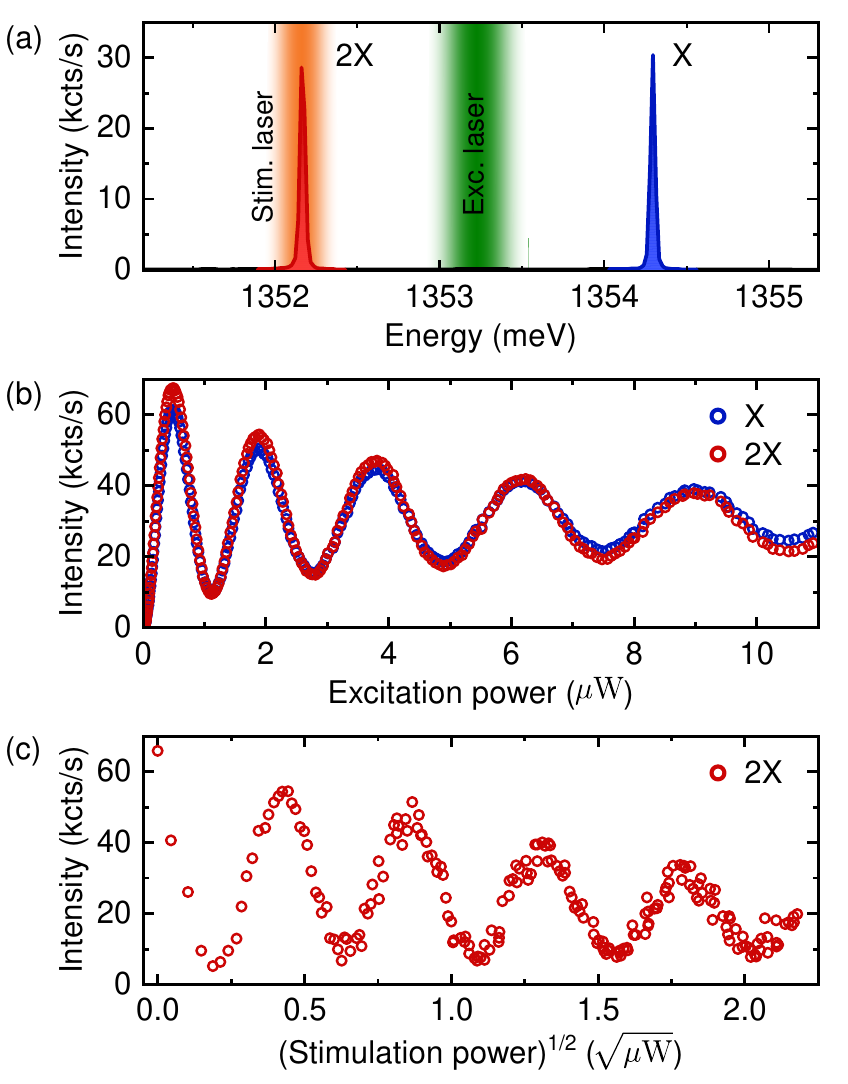}
    \caption{(a) Spectrum of the quantum dot emission after two-photon excitation of the biexciton. The biexciton (red) is red detuned from the neutral exciton transition (blue) by the biexciton binding energy $E_b$. The energy of the excitation and stimulation laser are depicted in green and orange, respectively. (b) Rabi oscillations between the ground and the biexciton state observed via the oscillation of the biexciton (red) and exciton (blue) integrated emission intensity as a function of the excitation laser power. (c) Integrated biexciton emission intensity as a function of the square root of the stimulation laser power. The system is coherently driven between the biexciton and exciton state.}
    \label{fig:figure2}
\end{figure}

We continue to experimentally confirm these findings using a semiconductor InGaAs quantum dot. 
A typical emission spectrum for the two-photon excitation of the $\ket{0} \leftrightarrow \ket{\textrm{2X}}$ transition is presented in Fig.~\ref{fig:figure2}(a).
The emission from both the biexciton and exciton is clearly visible with comparable intensity while the spectrally separated excitation laser is suppressed by cross-polarized filtering.
Rabi oscillations of the $\ket{0} \leftrightarrow \ket{\textrm{2X}}$ transition driven by a ${\sim}\SI{5.3}{\pico\second}$ long Gaussian pulse and measured via the biexciton and subsequent exciton emission (Fig.~\ref{fig:figure2}(b)) confirm the coherent nature of the two-photon excitation.
With increasing power up to a pulse area of $10\pi$, increasing damping is observed which is mainly caused  by coupling to LA phonons during the excitation process \cite{Ramsay2010}, as well as renormalization of the Rabi frequency that causes deviation from the expected linear behavior with increasing power \cite{Ramsay2010a}.
After the preparation of the system in the $\ket{\mathrm{2X}}$ state and a subsequent delay of ${\sim}\SI{9}{\pico\second}$ the stimulation pulse in resonance with the $\ket{\mathrm{2X}} \leftrightarrow \ket{\mathrm{X}}$ transition is applied to drive the system coherently between those states. Measuring the intensity of the biexciton emission as a function of the stimulation pulse power (Fig.~\ref{fig:figure2}(c)) reveals clean Rabi oscillations with an asymmetric damping (see Supplemental Material for details~\cite{Supplemental}).
Note that, for a pulse area of $\pi$, the biexciton emission is almost entirely suppressed indicating an efficient transfer of the population to the exciton state. For the following measurements we set the pulse area to $\pi$ for the excitation and stimulation laser.
\begin{figure}
    \centering
    \includegraphics[width=\linewidth]{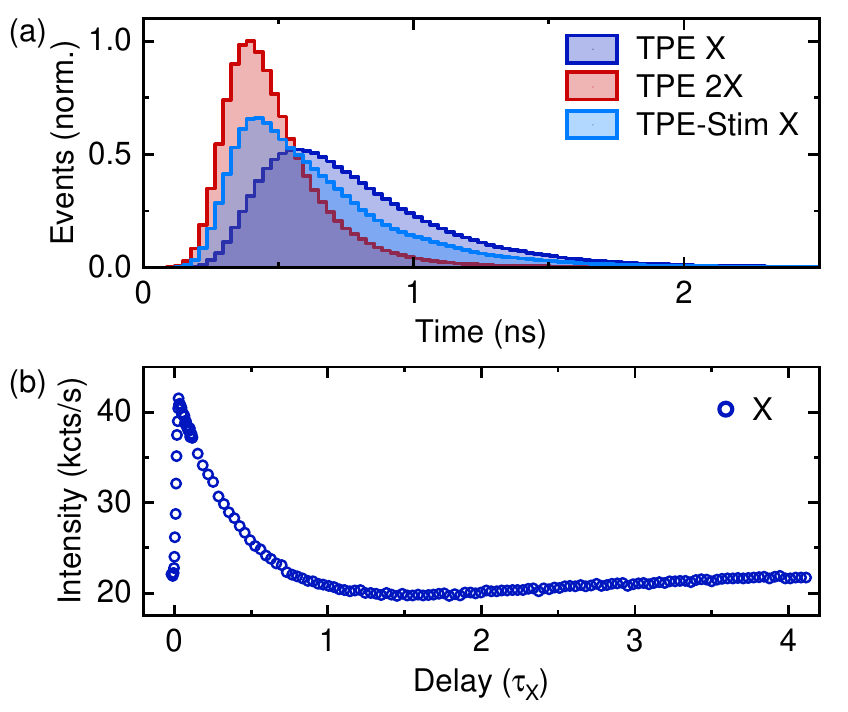}
    \caption{(a) Time-resolved fluorescence measurement of the quantum dot emission. The exciton decay (blue) is delayed with respect to the biexciton (red). Adding a stimulation pulse shortly after the biexciton preparation eliminates that delay (light blue). (b) Exciton emission intensity as a function of the stimulation pulse delay.}
    \label{fig:figure3}
\end{figure}

To gain insight into the system dynamics we perform time-resolved photoluminescence measurements of the two transitions subject to $\pi$-pulse excitation of the biexciton.
The obtained histograms are presented in Fig.~\ref{fig:figure3}(a) with each curve normalized to the sum of its events.
The biexciton directly decays after excitation with a monoexponential decay (red line) with a lifetime of ${\sim}\SI{179}{\pico\second}$.
Due to the cascaded emission, the population of the $\ket{\mathrm{X}}$ state has to build up before radiative recombination can occur. This leads to a delayed decay of the exciton emission (blue line).
Taking this into account we extract an exciton lifetime of ${\sim}\SI{293}{\pico\second}$.
By applying the stimulation $\pi$-pulse resonant on the $\ket{\textrm{2X}} \leftrightarrow \ket{\textrm{X}}$ transition shortly after the biexciton has been prepared, the exciton is instantaneously populated followed by its decay through spontaneous recombination (light blue line).

While resonant excitation usually requires cross-polarized filtering in the optical setup to distinguish between excitation laser and the generated single photons, the two-photon excitation allows simple spectral filtering.
This advantage persists when adding the stimulation laser as it is even further detuned from the transition of interest.
As already mentioned, the stimulation laser enables coherent control over the prepared fine structure split exciton state via polarization selectivity and therefore allows controlling the polarization of the emitted photon (H/V).
We demonstrate this by varying the delay of the stimulation pulse with respect to the excitation pulse and recording the exciton emission intensity shown in Fig.~\ref{fig:figure3}(b) using a co-polarized filtered setup.
If the stimulation pulse arrives before the biexciton is prepared it does not affect the population of the quantum dot states since it is far red-detuned from the exciton and two-photon resonance and phonon-assisted excitation processes have a negligible impact~\cite{Ardelt2014,Quilter2015}.
As a consequence, both decay paths and thus polarizations of the emitted photons have an equal probability leading to the detection of only half of the collected photons due to the polarization filter in the detection path.
Increasing the delay of the stimulation pulse increases the number of detectable photons with a maximum at ${\sim}\SI{9}{\pico\second}$ ($0.03\:\tau_{\mathrm{X}}$) delay where the measured intensity is almost twice as high as without the stimulation laser, consequently overcoming the limitation given by cross-polarized resonance fluorescence.
Further increasing the delay reduces the advantageous effect of the stimulation laser since the biexciton has had a higher chance to decay spontaneously.

\begin{figure*}
    \centering
    \includegraphics[width=\textwidth]{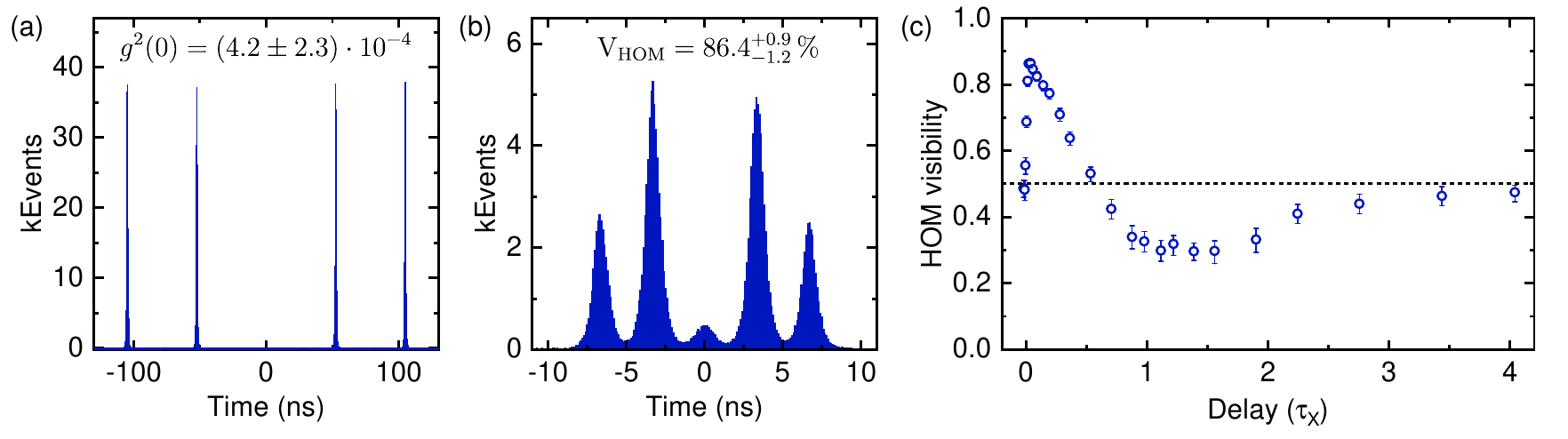}
    \caption{(a) Autocorrelation histogram of the exciton emission after stimulated preparation. (b) Correlation histogram of the exciton emission measured with an unbalanced Mach-Zehnder interferometer. (c) HOM visibility of subsequent exciton emission as a function of the stimulation pulse delay. The dashed line marks the HOM visibility without the stimulation pulse.}
    \label{fig:figure4}
\end{figure*}

As discussed above, two-photon excitation can be used to realize a near perfect single-photon source in terms of purity \cite{Schweickert2018,Hanschke2018}, but the indistinguishability of the emitted photons is intrinsically limited by the nature of the cascaded decay \cite{Scholl2020}.
Adding the stimulation pulse does not affect the single-photon purity if the two pulses do not overlap in time.
We demonstrate this by measuring the second-order coherence of the emitted exciton photons after the stimulated population transfer from the biexciton (Fig.~\ref{fig:figure4}(a)). We obtain $g^{(2)}(0) =(4.2\pm2.3)\cdot 10^{-4}$ with a purity that is mainly limited by a small background from imperfect laser suppression and dark counts of the detectors.
Finally, we study the indistinguishability of subsequently emitted photons by measuring the Hong-Ou-Mandel (HOM) visibility using an unbalanced Mach-Zehnder interferometer (Fig.~\ref{fig:figure4}(b))~\cite{Santori2002}. We observe a maximum value of $\textrm{V}_{\textrm{HOM}}=86.4^{+0.9}_{-1.2}\%$ for a time delay of ${\sim}\SI{9}{\pico\second}$, in good agreement with the value obtained under resonant excitation for the quantum dot under investigation (see Supplemental Material~\cite{Supplemental}). This observation suggests that the deviation from the theoretically predicted higher value results from a limited spectral stability of the studied sample~\cite{Thoma2016}.
Importantly, varying the time delay between the excitation and stimulation pulse, as shown in Fig.~\ref{fig:figure4}(c), reproduces the same behavior of the HOM visibility as predicted by our simulation.
If the stimulation pulse arrives before the system has been driven to the $\ket{\textrm{2X}}$ state we observe the same HOM visibility as without the stimulation pulse (dashed line).
Introducing a short delay (${\sim}\SI{9}{\pico\second}$) yields the maximum indistinguishability followed by a fast decline even lower than for bare two-photon excitation with a recovery of the value for delays larger than $3\tau_{\mathrm{X}}$. 
Note that the experimentally measured visibility over the entire measurement range is slightly lower compared to the simulations which arises from neglecting dephasing in the model (see Supplemental Material for details~\cite{Supplemental}).

In summary, we proposed a scheme for the generation of single photons based on two-photon excitation of the biexciton followed by a stimulation of the biexciton to exciton transition and demonstrated its potential both experimentally and in numerical simulations.
Since both laser pulses are detuned from the single-photon emission spectral filtering is sufficient to separate emission and excitation which enables high brightness in comparison to polarization filtering.
Moreover, the polarization-selective coupling of the stimulation pulse to the fine structure split exciton states enables deterministic programming of the polarization of emitted photons (H/V) in real-time.
Since this scheme combines the advantages of previously demonstrated excitation methods we expect it to play a significant role in the future development of high-fidelity deterministic single-photon sources and their applications in photonic quantum technologies.

We gratefully acknowledge financial support from the German Federal Ministry of Education and Research via the funding program Photonics Research Germany (Contract No. 13N14846), the European Union's Horizon 2020 research and innovation program under Grants Agreement No. 820423 (S2QUIP), No. 862035 (QLUSTER) and No. 899814 (Qurope), and the Deutsche Forschungsgemeinschaft (DFG, German Research Foundation) via SQAM (F-947-5/1) and Germany's Excellence Strategy (MCQST, EXC-2111, 390814868).

\providecommand{\noopsort}[1]{}\providecommand{\singleletter}[1]{#1}%

\newpage

\end{document}